\documentclass[11pt]{article}

\usepackage{a4wide}
\usepackage{xspace}
\usepackage{multicol}
\usepackage{amsmath}
\usepackage{graphicx}
\usepackage{epsfig}
\usepackage{hyperref}
\usepackage{url}
\usepackage{amssymb}

\newcommand{\be}{\begin{equation}}
\newcommand{\ee}{\end{equation}}

\newcommand{\bea}{\begin{eqnarray}}
\newcommand{\eea}{\end{eqnarray}}

\def\order#1{{\cal{O}}\left(#1\right)}

\newcommand{\hard}{\text{hard}}

\newcommand{\GeV}{\,\mathrm{GeV}}
\newcommand{\TeV}{\,\mathrm{TeV}}

\title{\textbf{Jet Fragmentation Function Moments\\
    in Heavy Ion Collisions}}
\author{
  Matteo Cacciari,\!$^{1,2}$ Paloma Quiroga-Arias,$^1$ Gavin P.~Salam$^{3,4,1}$ and Gregory Soyez$^{5}$ \\
\\
{\sl  \small $^1$LPTHE, UPMC Univ.~Paris 6 and CNRS UMR 7589, Paris, France}\\[2pt]
{\sl  \small $^2$Universit\'e Paris Diderot, Paris, France}\\[2pt]
{\sl  \small $^3$CERN, Department of Physics, Theory Unit, CH-1211 Geneva 23, Switzerland}\\[2pt]
{\sl  \small $^4$Department of Physics, Princeton University, Princeton, NJ 08544, USA}\\[2pt]
{\sl  \small $^5$Institut de Physique Th\'eorique, CEA Saclay, CNRS URA 2306, F-91191 Gif-sur-Yvette, France}
}
\date{}

\begin{document}

\maketitle

\vspace{-9.5cm}
 \begin{flushright}
   CERN-PH-TH/2012-227\\
   September 2012\\
 \end{flushright}
\vspace{8cm}

\begin{abstract}
  The nature of a jet's fragmentation in heavy-ion collisions has the
  potential to cast light on the mechanism of jet quenching.
  However the presence of the huge underlying event complicates the
  reconstruction of the jet fragmentation function as a function of the
  momentum fraction $z$ of hadrons in the jet.
  Here we propose the use of moments of the fragmentation function.
  These quantities appear to be as sensitive to quenching
  modifications as the fragmentation function directly in $z$.
  We show that they are amenable to background subtraction using the same
  jet-area based techniques proposed in the past for jet $p_t$'s.
  Furthermore, complications due to correlations between
  background-fluctuation contributions to the jet's $p_t$ and to its
  particle content are easily corrected for.
\end{abstract}

\clearpage

\tableofcontents
\newpage

%
\section{Introduction}\label{sec:intro}

In heavy-ion physics one is interested in studying the hot and dense
medium that is formed by the high energy collisions of heavy
nuclei. One way of accessing such properties is to look at how the
characteristics of a number of `hard probes' are modified by their
passage through this medium.  
While early examples of hard probes consisted of quarkonia or single
high transverse-momentum ($p_t$) hadrons, the performance and the
acceptance of the Large Hadron Collider (LHC) detectors (ALICE, ATLAS
and CMS) allow for the in-depth study of fully-fledged jets.
A number of experimental results on jets in heavy ion collisions have
already been published following the first two heavy ion runs in 2010
and 2011. 
Of particular interest was the observation by ATLAS~\cite{Aad:2010bu}
and CMS~\cite{Chatrchyan:2011sx, Chatrchyan:2012nia} of a sizeable
asymmetry in the transverse momentum of dijet pairs, which was
interpreted (modulo some caveats~\cite{Cacciari:2011tm} concerning a
possible partial spoofing role of background fluctuations) as evidence
of some amount of quenching, i.e.\ of a reduction of a jet's momentum
due to its interaction with the medium.
A better understanding, also at the
quantitative level, of this phenomenon would naturally allow one to
better constrain the characteristics of the medium itself.

After studying an inclusive observable like the quenching of a full
jet, the next logical step is to observe the momentum distribution of
its constituents, i.e.\ its fragmentation function (FF). One would
like to ascertain if and how the FF is modified, due to the
interactions with the medium, with respect to the FF of jets produced
in proton-proton collisions and fragmenting instead in the vacuum.

Measurements of jet fragmentation functions have been performed since
the first observations of hadronic jets. 
LEP \cite{Buskulic:1995aw,Abreu:1997ir,Abbiendi:2004pr} and
Tevatron~\cite{Abe:1990,Acosta:2002gg} 
experiments studied them in detail in $e^+e^-$ and $p\bar
p$ collisions respectively and compared them to accurate calculations
in perturbative QCD, finding very good agreement. 
LHC experiments \cite{Aad:2011sc,Chatrchyan:2012gw} more recently
measured them in $pp$ collisions.

In a heavy ion context, theoretical discussion as to how the
fragmentation function could be expected to be modified by the medium
were put forward, for instance, in
\cite{Borghini:2005em,Guo:2000nz,Armesto:2007dt,Arleo:2008dn,Sapeta:2007ad,Majumder:2009zu,Beraudo:2011bh},
and the question can be examined also with the help of Monte Carlo
programs that simulate jet fragmentation in a
medium~\cite{pyquen,Zapp:2008gi,Renk:2008pp,Armesto:2009fj,Schenke:2009gb}.
The study of jet FFs in a heavy-ion context is however complicated by
the presence of a large and predominantly soft background due to the
underlying event produced by the nucleons not directly participating
in the hard interaction.
The background acts in essentially two ways: it adds many soft
particles to the jet --- an order of magnitude more than are normally
present in a jet --- severely contaminating the fragmentation function
at low hadron momentum fractions $z$; and it can bias the measured
$p_t$ of the jet, even after subtracting the expected average $p_t$
shift, skewing the value one reconstructs for a given hadron's
momentum fraction.

Experimentally, there have so far been measurements of jet FFs by three
experiments, in AuAu collisions at RHIC by the STAR collaboration
\cite{Putschke:2008wn,Bruna:2009em}, and more recently in PbPb
collisions at the LHC by the
CMS~\cite{Chatrchyan:2012gw,CMS-PAS-HIN-12-013} and
ATLAS~\cite{etzion-ichep,ATLAS-CONF-2012-115} collaborations.
In each case, the measurements were limited to hadron $p_t$'s above
$1-4$~GeV, where soft contamination is less important.
Different approaches were used to address the biases in the
determination of $z$, for example unfolding this effect in the case of
ATLAS, or folding the effect into the $pp$ reference FF in the case of
CMS, the only experiment which goes down to $1\GeV$.
The largest changes of heavy-ion FFs relative to reference $pp$ FFs
appear to be in the region of small $z$, where uncertainties
associated with the background subtraction grow.
It would clearly be of interest to probe the fragmentation functions
at lower $z$ and with higher accuracy, especially as other
measurements also indicate that soft particles appear to play an
important role in quenching~\cite{Chatrchyan:2011sx}.

The purpose of this article is to suggest the use of a new observable
that may help provide an alternative way of understanding
fragmentation patterns in heavy-ion jets: momentum-fraction weighted
moments of the fragmentation functions.

Moments are often used in theoretical calculations, and have been helpful
in the past in diagnosing and resolving experiment-theory
discrepancies~\cite{Cacciari:2002pa}, but to our knowledge have not
been examined so far in heavy-ion collisions.
Among their benefits, one can note that they are helpful in directly
relating jet spectra with hadron spectra and also that they are
sensible objects on a jet-by-jet basis (whereas event-by-event,
fragmentation functions are simply sums of individual delta-functions,
one for each hadron in the jet).
The latter point makes it feasible to measure correlations within an
event between any given moment of the background fragmentation
spectrum and fluctuations in the background's $p_t$.
As we shall see, it seems that this can be of significant benefit in
extracting the soft part of the jet's fragmentation and also in
reducing systematics in the extraction of the hard part.

%
\section{Representations of the fragmentation
  function}\label{sec:representations}

A jet fragmentation function can be defined as the distribution
$dN_{h}/dz$ of the momentum fraction
\begin{equation}
  \label{eq:z}
  z = \frac{p_{t,h}}{p_t} \, ,
\end{equation}
of hadrons in the jet, where $p_{t,h}$ is the transverse momentum of
the hadron and $p_t$ is the transverse momentum of the
jet.\footnote{Various other definitions are also in use, for example
  replacing transverse momenta with 3-momenta and taking the
  projection of the hadron momenta along the axis of the jet.}

\begin{figure}[t]
\centerline{\includegraphics[angle=270,width=\textwidth]{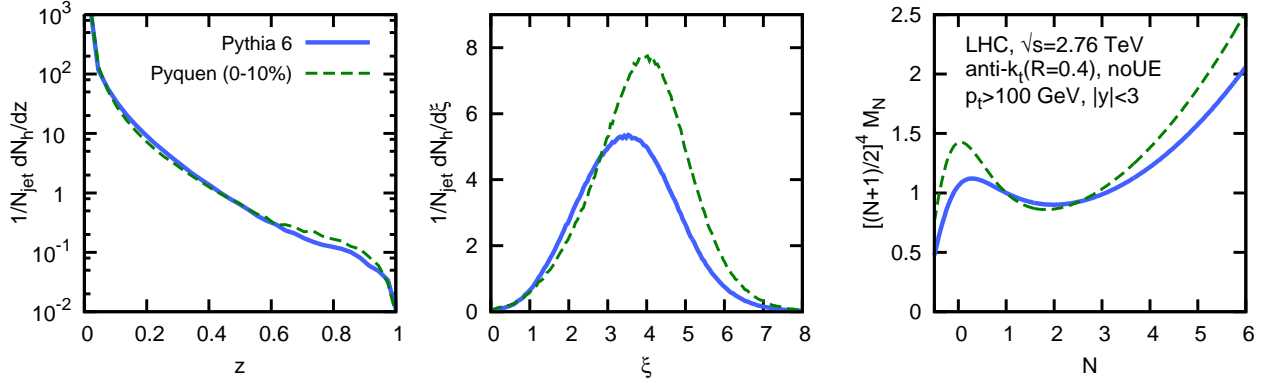}}
\caption{Jet fragmentation functions (versus $z$ and $\xi$) and moments
  (versus $N$) in proton-proton collisions at the 
  LHC ($\sqrt{s_{NN}} = 2.76$~TeV), obtained without quenching, from
  Pythia~6, and with quenching, from Pyquen.
}
\label{fig:hard}
\end{figure}

Figure~\ref{fig:hard} shows, in its left-most plot, the typical shape
of the $z$ distribution (normalised in this case to the total number
of jets $N_{\text{jet}}$ used in the analysis), given for anti-$k_t$
($R=0.4$) jets~\cite{antikt} with $p_t > 100\GeV$ in $pp$ collisions
at a centre-of-mass energy of $2.76\TeV$.
The middle plot shows identical information, represented as a
differential distribution in $\xi = \ln 1/z$.
The $z$ representation helps visualise the hard region of the FF, while
$\xi$ devotes more space to the soft part.
Two curves are shown: one of them, solid blue, labelled ``Pythia 6'', is a $pp$
reference curve obtained from Pythia 6.425~\cite{pythia} with its
virtuality-ordered shower.
The other, dashed green, was obtained with the Pyquen program~\cite{pyquen,
  pyquen_tune}, which modifies Pythia showering so as to simulate
quenching. We have used it with settings corresponding to $0-10\%$
centrality.
Its effect on the FF is of the same order of magnitude as the effects
seen experimentally~\cite{CMS-PAS-HIN-12-013,ATLAS-CONF-2012-115} and it 
therefore provides a useful reference when
establishing whether FFs are being reconstructed with sufficient
accuracy by some given procedure.

The right-most plot of Fig.~\ref{fig:hard} shows moments of the
fragmentation functions.
The $N^\text{th}$ moment, $M_N$, of the fragmentation function is
given by the integral
\begin{equation}
  M_N = \frac{1}{N_{\text{jet}}}\int_0^1 z^N\, \frac{dN_{h}}{dz}\, dz
      = \frac{1}{N_{\text{jet}}}\int_{0}^{\infty} e^{-N \xi}\, \frac{dN_{h}}{d\xi}\, d\xi\,.
\end{equation}
In practice, the moments for a single jet can be calculated as
\begin{equation}
  \label{eq:MN-direct}
  M_N^{\text{jet}} =\frac{\sum_{i\in{\text{jet}}} p_{t,i}^N}{p_t^N} \, ,
\end{equation}
where the sum runs over all the jet's constituents. The results can
then be averaged over many jets, so that $M_N = \left\langle
  \smash{M_N^{\text{jet}}}\right\rangle_{\text{jets}}$.  Obviously, $M_0$ represents
the average particle multiplicity in a jet, and $M_1$ is equal to one
by virtue of momentum conservation if one measures all hadrons, as we
assume here (taking $\pi_0$'s to be stable).\footnote{\label{foot} The relation $M_1 = 1$ is exact
  only if one defines the transverse momentum of a jet as the scalar
  sum of the transverse momenta of the constituents, so that the
  moments are given by
  \begin{equation}
    \label{eq:MN-exact}
    M_N^{\text{jet}} =\frac{\sum_i p_{t,i}^N}{\left(\sum_i p_{t,i}\right)^N} \,.
  \end{equation}
  This is the choice that has been made in this paper. Using the
  transverse component of the jet momentum rather than the scalar sum
  leads to small violations (of the order of a fraction of one percent, 
  for the jet radius $R=0.4$ used in this work) of
  the $M_1=1$ relation.
  If the fragmentation moments are measured with only charged hadrons,
  we will still assume that the denominator of
  Eq.~(\ref{eq:MN-direct}) is determined with all particles in the
  jet or, equivalently, with all calorimeter towers.
}
If instead only charged tracks are used in the numerator, then it is
clear that $M_1$ will be significantly below $1$.
There is another value of $N$ that is of special interest: given a jet
spectrum $d\sigma_\text{jet}/dp_t$ that falls as $p_t^{-n}$, the ratio
of the inclusive hadron spectrum and inclusive jet spectrum is given
by $M_{n-1}$.
Thus, $M_{n-1}^\text{AA}/M_{n-1}^\text{pp}$ corresponds to the ratio
of (charged-)hadron and jet $R_\text{AA}$ values (in the approximation that
$n$ is exactly independent of $p_t$),
\begin{equation}
\frac{M_{n-1}^{\text{AA}}}{M_{n-1}^{\text{pp}}}= \frac{R_{\text{AA}}^{\text{h}}}{R_{\text{AA}}^{\text{jet}}}\, .
\end{equation}
For $p_{t}$ in the range $100-200\GeV$, at $\sqrt{s}_{NN}=2.76\TeV$,
the relevant $n$ value has some dependence on $p_t$ and is in the
range $n=6-7$, corresponding to $N=5-6$.

\begin{figure}[t]
\centerline{\includegraphics[angle=270,width=0.4\textwidth]{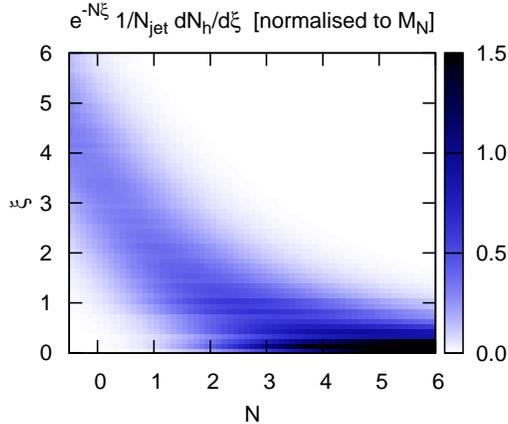}}
\caption{Representation of the $\xi$ values that contribute dominantly
  to the $M_N$ integral for a given $N$, shown as a function of
  $N$. Shown for the Pythia~6 results and cuts of
  Fig.~\ref{fig:hard}.}
\label{fig:ximap}
\end{figure}

In representing the moments in figure~\ref{fig:hard}(right), we
include a factor $((N+1)/2)^4$, which allows a broad range of $N$
values to be shown on a linear vertical scale.
The same features visible in the plots versus $z$ and $\xi$ are
visible versus $N$ too, for example that Pyquen leads to higher
multiplicities than Pythia at small $N$ (corresponding to $z < 0.05$
or $\xi > 3$)
and large $N$ ($z>0.5$, $\xi < 0.7$), and a slightly reduced multiplicity at
intermediate $N$ ($z\sim 0.2$, $\xi \sim 1.5$).

To help understand the quantitative relationship between $N$ and
$\xi$, one may examine figure~\ref{fig:ximap}, a colour-map that shows
as a function of $N$ the contribution to the $M_N$ moment from each
$\xi$ value.
It shows clearly how large $\xi$ values dominate for low $N$ (and vice-versa).
This $\xi, N$ relationship depends to some extent on the shape of the
fragmentation function and it is given for the same Pythia~6
fragmentation function that was used in Fig.~\ref{fig:hard}.

\section{Simulation and reconstruction tools}

Before proceeding with the heavy-ion analysis, it is useful to detail
the simulation tools that we use. 

What we call `hard' jets in QCD are simulated in proton-proton ($pp$)
collisions using Pythia~6.425~\cite{pythia} in dijet mode with the DW
tune~\cite{Albrow:2006rt}.\footnote{This tune is not the most up to
  date; however it is not unrealistic for the LHC and has the
  characteristic that it is based on Pythia's virtuality ordered
  shower, which is a prerequisite for use with Pyquen.}
We have not included any $pp$ underlying event (UE), its effect being
minimal in this context anyway.
Jets including quenching effects have been generated using
Pyquen~\cite{pyquen, pyquen_tune}, v1.5.1, as included in
Hydjet~\cite{Lokhtin:2003ru,hydjet}.
We consider all hadrons, not just charged tracks, and take $\pi_0$'s
to be stable, so that we are still considering genuine hadron
distributions.

The heavy-ion background (which we will also call heavy-ion underlying
event) is simulated using Hydjet v1.6 for
0-10\% centrality and, where needed, it is superimposed to the hard
event generated by Pythia.\footnote{In their notes, CMS and ATLAS often call this procedure 'embedding' or 'overlying' respectively.}
The jets observed in this combined event will be denoted as `full'
jets. 

All events are generated for either $pp$ or lead-lead (PbPb)
collisions at the LHC, with a centre-of-mass energy of 2.76 TeV per
nucleon-nucleon collision.

Jets are reconstructed using the anti-$k_t$ algorithm~\cite{antikt} with
$R=0.4$, as implemented in FastJet~\cite{fastjet,fastjet_fast}.
In estimating the jet's $p_t$, the HI background is subtracted using
the median/area based techniques introduced
in~\cite{subtraction,areas,hipaper}, which we implement with $k_t$
jets~\cite{kt} ($R=0.4$) to estimate $\rho$, the background transverse
momentum per unit area.
At most the two hardest jets passing a hard cut on the subtracted
transverse momentum (see below) are subsequently used for the fragmentation
function analysis.
The cuts that we shall use are $100\GeV$ and $200\GeV$.
%

\section{Impact of HI background and its subtraction}
\label{sec:subtraction}

As explained in the Introduction, the addition of the heavy-ion background has the potential to modify
a measured fragmentation function in two ways.
Firstly, the jet's $p_t$ is modified, affecting the normalisation of
$z$ in Eq.~(\ref{eq:z}).
Secondly, the heavy-ion background adds many extra particles to the
jet, predominantly at low momenta.

To appreciate the impact of the extra particles in the jet from the
heavy-ion background, it is instructive to first examine the 
``Pythia+Hydjet'' dashed red curves of Fig.~\ref{fig:hydjet}.
These show the FF extracted in heavy-ion collisions, without any
subtraction of the background contribution to the FF,
but always using a $z$ value defined such that the jet's $p_t$ has
been corrected for the expected HI background contamination (as is
standard in the experimental measurements).
At this stage we will not perform any unfolding to account for
fluctuations in the HI background.

One sees how the FF acquires a `bump' in the soft
region, which lies at larger $\xi$ (smaller $z$) than the maximum in the original
$pp$ result (blue solid line) and is up to two orders of magnitude
higher. 
For the moments, this presence of the background is seen as a steep
increase in the small-$N$ region, taking the curve far off the scale.

At large $z$ (small $\xi$), the impact of the addition of the
background is only barely visible, as might be expected given that it
is dominated by soft particles.
However, this visually small effect is partially an artefact of the
logarithmic scale used to show the FF versus $z$.
Considering instead the moments, one sees that there is a
non-negligible \emph{reduction} in the FF at large $N$.
This is perhaps surprising given that the background adds particles.
It is a feature related to the interplay between background fluctuations
and the steeply falling jet spectrum. It is well known by the
experiments and we will return to discuss it in
section~\ref{sec:improved}.

\subsection{$z$-space subtraction}

\begin{figure}[t]
  \begin{center}
    \includegraphics[angle=270,width=0.90\textwidth]{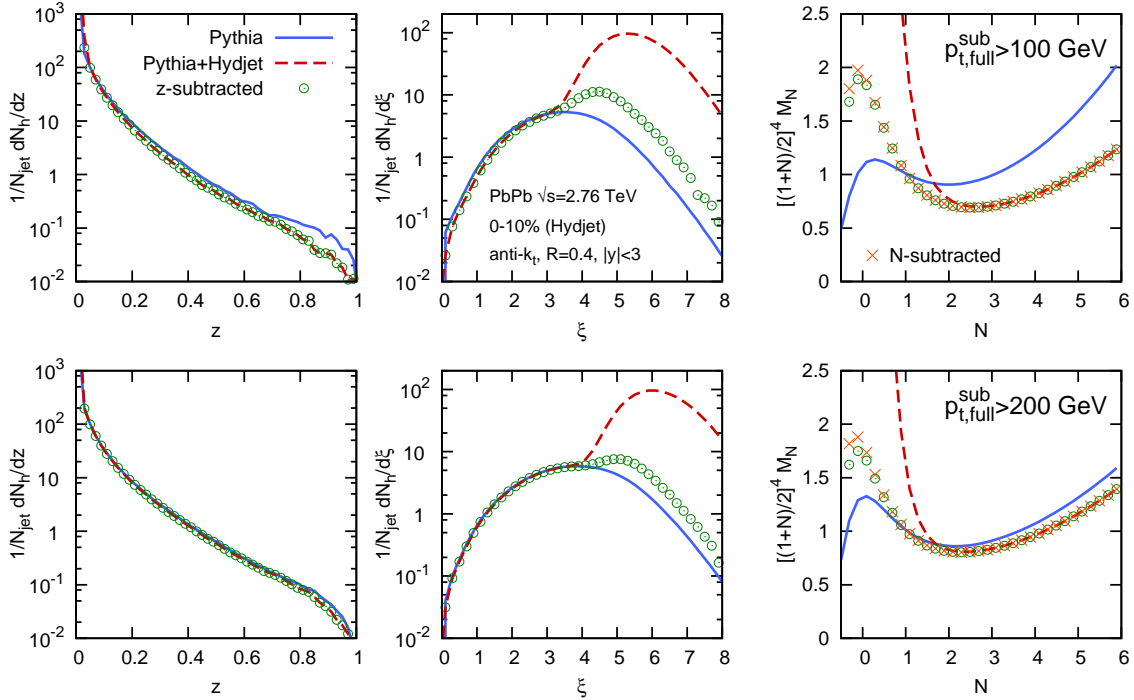}
    \caption{Jet fragmentation functions shown for plain Pythia, with
      the addition of the heavy-ion background (Pythia+Hydjet) and
      after subtraction of the heavy-ion background ($z$-subtracted
      and $N$-subtracted).
      For the results including the heavy-ion background, the jet
      $p_t$ used to define $z$ is always that after subtraction of the
      heavy-ion background. 
      As in Fig.~\ref{fig:hard}, we show the results as a function of
      $z$ (left), $\xi$ (middle) and for the moments versus $N$ (right).
      The upper (lower) row has a jet $p_t$ threshold of $100\GeV$
      ($200\GeV$). 
      \label{fig:hydjet}}
  \end{center}
\end{figure}

The traditional approach to background subtraction from a jet FF
involves the construction of a distribution (in $z$ or $\xi$ space) intended
to approximate that of background-only particles, and the
subtraction of this distribution from the measured one. 
The way the background-only distribution is determined can vary,
the simplest one probably being to consider a region of the event
that is expected to be little affected by the hard jets, and
measure it there. 

To illustrate $z$-space subtraction here, we measure the distribution
of hadrons in two regions transverse in azimuth with respect to the
axis defined by the dijet event.
Event by event, and jet by jet, we subtract those distributions
(measured in a patch of phase space with an area equal to that of the
jet that one is considering) from the jet fragmentation function.
While exact experimental procedures differ in the details, most
choices lead to similar results here.
Perhaps the main distinction of the experimental procedures is that
they sometimes address issues related to flow, which for simplicity we
neglect here in our $z$-space subtraction.

The results of this subtraction are shown in figure~\ref{fig:hydjet}
as green open circles, labelled `$z$-subtracted' (the moments of these
$z$-subtracted results are also shown).
One sees how these curves come closer to the $pp$ results at small $z$
and small $N$ than do the unsubtracted (dashed red) ones.
The agreement improves as the jet $p_t$ threshold is increased.
However, even with jets of $p_t \gtrsim 200$~GeV, this procedure still
falls short of an accurate reconstruction of the hard FF. 
In fact, the region in $z$ where the subtracted FF works well barely
extends beyond that selected by simply truncating the unsubtracted FF
at low $p_{t,\text{h}}$ so as to avoid the region dominated by the
background.
 
\subsection{$N$-space subtraction}

An alternative approach is to directly subtract the moments $M_N$ of
the fragmentation function. 
This can be done by extending the jet-area based techniques for
background estimation introduced in \cite{subtraction} and implemented
in \cite{fastjet}. 

Let us first recall the procedure for subtracting the background
contribution to a jet's $p_t$.
One first determines the background $p_t$ density per unit
area, $\rho$, for the event (or just part of the event in the vicinity of
the jet). 
One accomplishes this by dividing the event into patches of similar
rapidity-azimuth area, e.g.\ by running the inclusive $k_t$ algorithm
and taking all the jets it finds.
Then $\rho$ is obtained by taking the median across all patches of
the ratio of $p_t$ to area ($A_\text{patch}$) for each patch:
\begin{equation}
  \rho =
  \mathop{\mathrm{median}}_\text{patches}\left\{\frac{p_{t,\text{patch}}}{A_\text{patch}}\right\}\,.
\end{equation}
The median serves to limit biases from any hard jets among the patches.
A given jet's full transverse momentum, $p_{t,\text{full}}$, which
includes genuine jet particles and background particles, is then
corrected for the background particles by subtracting an amount given
by the product of $\rho$ and the jet's area $A$:
\begin{equation}
\label{eq:ptsub}
  p_{t,\text{full}}^\text{sub} = p_{t,\text{full}} - \rho A\,.
\end{equation}
We use patches in an annulus (or `doughnut') of outer radius $3R$
and inner radius $R$, centred on the jet of interest.
This choice helps to limit corrections due to flow and accounts for
the rapidity dependence of the background.\footnote{Alternatively one
  could determine a global $\rho$ for the event and modulate it for
  rapidity and azimuthal dependence, as discussed in \cite{fastjet},
  and similarly to what is done for example by
  ATLAS~\cite{ATLAS:2012is}.}

For the purpose of correcting FF moments, the procedure we propose is
quite similar.
One first determines the expected background contribution per unit
area to a given moment (or rather to its numerator in
Eq.~(\ref{eq:MN-direct})):
\begin{equation}
  \rho_N =
  \mathop{\text{median}}_\text{patches}\left\{\frac{\sum_{i\in \text{patch}}
  p_{t,i}^N}{A_\text{patch}}\right\}\,.
\end{equation}
For example, for $N=0$ this will be the median particle multiplicity
per unit area.
The subtracted FF moment is then obtained by separately taking the
numerator and denominator of Eq.~(\ref{eq:MN-direct}), as measured in
the full event, and respectively subtracting $\rho_N A$ and $\rho A$:
\begin{equation}
  \label{eq:momsub}
  M_N^\text{sub} =\frac{\sum_{i\in \text{jet}} p_{t,i}^N - \rho_N A}{(p_{t,\text{full}} - \rho
    A)^N} \, .
\end{equation}

The results for $M_N^\text{sub}$ are shown in figure~\ref{fig:hydjet} in
the right-most plots, orange crosses labelled `$N$-subtracted'. One
can see that they are neither better nor worse than the corresponding
$z$-subtracted ones, which have been translated to $N$-space and drawn
in the same plots for direct comparison (green circles).

In order to provide a fragmentation function reconstruction that is
better than that given by the standard $z$-space method, we
introduce in the next section an improved background subtraction
method that is most straightforwardly applied in moment space.

%
\section{Improved background-subtracted fragmentation
  function}\label{sec:improved}

The discrepancies that we have observed can, we believe, largely be
attributed to fluctuations of the background.

Background fluctuations mean that neither the jet's $p_t$ nor the
numerator of the FF moment are perfectly reconstructed for any given
jet.
When selecting jets above some $p_t$ threshold for a process with a
steeply falling jet spectrum, it is favourable to select
jets slightly below the $p_t$ threshold, but which have an upwards
background fluctuation (cf.\ the discussion of section 5 of
\cite{hipaper}).
One consequence of this is that the denominator in Eq.~(\ref{eq:momsub})
tends to be larger than the actual jet $p_t$.
The larger the value of $N$, the greater the impact of this effect.
One can also understand it as resulting in an underestimate of the $z$
fraction, leading to the FF being spuriously shifted to lower $z$.
Such an effect is already well known: CMS~\cite{Chatrchyan:2012gw},
for example, when comparing with $pp$ FFs, explicitly applies a
correction to the $pp$ FFs to account for the smearing of the
denominator of the $z$ variable that is expected when the jet's
momentum is reconstructed in a HI environment.
ATLAS~\cite{ATLAS-CONF-2012-115} carries out an unfolding to
account for this.

A second consequence of fluctuations affects mostly the low-$z$, or
low-$N$ region of the FF: the upwards fluctuations of the background
$p_t$ that cause a jet to pass the $p_t$ cut even when it is below
threshold also tend to be associated with upwards fluctuations of the
multiplicity of soft background particles.
It is this effect that causes the FF to be overestimated for low
values of $N$.%
\footnote{As this letter was being finalised, ATLAS presented
  preliminary results~\cite{ATLAS-CONF-2012-115} in which they correct
  for this by multiplying the expected background contribution by
  a $z$-independent factor that is a function of the jet $p_t$.}

One way of verifying the above interpretation is to consider
$\gamma+$jet events, selecting events based on the photon $p_t$ and
normalising the fragmentation function $z$ also to the photon $p_t$.
In this case, the background-induced fluctuations of the jet's
reconstructed $p_t$ are of no relevance, and in explicit simulations,
we have found that plain subtraction is already quite effective.

For the dijet case, in the limit where the background fluctuations are
reasonably small compared to the transverse momentum of the jet, it is
possible to devise a simple correction in moment space for both kinds
of fluctuation-induced bias.
In order to do so, we start by rewriting the subtracted moments of the
jet FF in Eq.~(\ref{eq:momsub}) as
\begin{equation}
M_N^\text{sub} = \frac{\sum_i p_{t,i}^N - \rho_N A}{(p_{t,\text{full}} - \rho
A)^N} \equiv \frac{S_N}{S_1^N}
\end{equation}
where we have introduced the
shorthands $S_N$ for the subtracted numerator and $S_1$ for
$p^\text{sub}_{t,\text{full}}$, as defined in Eq.~\eqref{eq:ptsub}. 
Given our assumption that background fluctuations are moderate, we can
locally approximate the hard jet cross-section by an 
exponential rather than by a power in order to facilitate our
analytical working,
\begin{equation}
\label{eq:Hspectrum}
  H(p_t)\equiv \frac{d\sigma}{dp_t} = \frac{\sigma_0}{\mu}\,{\rm exp}(-p_t/\mu)\,.
\end{equation}
Next we take a Gaussian approximation for the spectrum of background
transverse-momentum fluctuations from one jet to the next.
Denoting the fluctuation by $q_t$, 
the distribution of $q_t$ is then
\begin{equation}
\label{eq:Bspectrum}
  B(q_t)\equiv \frac{dP}{dq_t} 
  = \frac{1}{\sqrt{2\pi A}\sigma}\,
       \exp\left(-\frac{q_t^2}{2\sigma^2 A}\right),
\end{equation}
where $\sigma$ is a parameter that describes the size of
fluctuations from one patch of area $1$ to another. 
It can be extracted directly from the event in a manner similar to
$\rho$~\cite{subtraction,css}. 
Its value is of the order of $\sigma \simeq 18$~GeV for our PbPb LHC 
($\sqrt{S_{NN}} = 2.76$~TeV)
simulations (the same simulation leads to fluctuations of $\sigma_\text{jet} \simeq 11\GeV$ for anti-$k_t$
$R=0.4$ charged-track jets, in
good agreement with the measurement from ALICE~\cite{Abelev:2012ej}).

We further introduce the variable $Q_N$ to denote the difference
between the actual background contribution to $S_N$ in a specific jet
and the expected contribution, $\rho_N A$, i.e.
\begin{equation}
  Q_N=\left(\sum_{i \, \in\, \text{jet\,(bkgd)}} k_{t,i}^N \right)-\rho_N A \, , 
\end{equation}
where the sum runs just over the background constituents $k_{t,i}$ of the
jet. By construction, $Q_1 = q_t$. 
In practice $Q_N$ cannot be determined for a single jet, since we
don't know which particles are the background ones, but its
statistical properties can be determined by looking at many jets.
The fluctuations of $Q_N$ are not independent of the
fluctuations $q_t$ of the background's transverse momentum: there is a correlation
coefficient $r_N$ between them, defined as
\begin{equation}
  r_N = \frac{\mathrm{Cov}(q_t,
    Q_N)}{\sqrt{\mathrm{Var}(q_t)\mathrm{Var}(Q_N)}}\,,
\end{equation}
where $\mathrm{Var}(X)$ is the variance of the variable $X$ and
$\mathrm{Cov}(X,Y)$ the covariance of $X$ and $Y$.
Using the fact that $\mathrm{Var}(q_t) = \sigma^2 A$,
$\mathrm{Var}(Q_N) = \sigma_N^2 A$, we have that the average value for
$Q_N$ as a function of $q_t$ is
\begin{equation}
  \label{eq:expected-QN}
  \langle Q_N \rangle(q_t) 
  = \frac{\mathrm{Cov}(q_t, Q_N)}{\mathrm{Var}(q_t)} q_t
  = r_N \frac{\sigma_N}{\sigma} q_t\,.
\end{equation}

With these ingredients we can now correct for the
fluctuation effects as follows.
Firstly, we note that the cross section for a given reconstructed
$S_1$ (i.e.\ subtracted jet $p_t$) is
\begin{equation}
  \label{eq:dsigma_dpt}
  \frac{d\sigma}{d S_1} = \int dq_t H(S_1 - q_t) B(q_t)\,,
\end{equation}
where the integral allows us to deduce the probability distribution
for the actual fluctuations $q_t$ given $S_1$:
\begin{equation}
  \label{eq:qt-dist}
  \left.\frac{dP}{dq_t}\right|_{S_1} = 
  \frac{H(S_1 - q_t) B(q_t) }{\int dq_t' H(S_1 - q_t') B(q_t')}
\end{equation}
corresponding, with our approximations for $H(p_t)$ and $B(q_t)$ in Eqs.~(\ref{eq:Hspectrum}) and 
(\ref{eq:Bspectrum}) respectively, to an
average $q_t$ of
\begin{equation}
  \label{eq:av-qt}
  \langle q_t \rangle = \frac{\sigma^2 A}{\mu}\,.
\end{equation}
If we measure a certain value $S_N$ in a jet, then as a function of the $q_t$
fluctuation, the expected true hard contribution to it is
\begin{equation}
  \label{eq:SN-hard-v-qt}
  S_N^\text{hard} = S_N - \langle Q_N\rangle(q_t)
   = S_N - r_N\frac{\sigma_N}{\sigma} q_t\,,
\end{equation}
where we have averaged over possible $Q_N$ values, given the
$q_t$ fluctuation.
To obtain our estimate for $M_N^\text{hard}$, this should be normalised by the
$N^\text{th}$ power of the true hard $p_t$ of the jet, $(S_1 -
q_t)^{N}$:
\begin{equation}
  \label{eq:MN-hard-v-qt}
  M_N^\text{hard} = \frac{S_N^{\hard}}{(S_1 - q_t)^{N}}
  = \frac{S_N}{S_1^N} + N \frac{S_N q_t}{S_1^{N+1}} 
         - r_N\frac{\sigma_N q_t}{\sigma S_1^N} + \order{q_t^2}\,.
\end{equation}
One subtlety here is that this is an estimate for $M_N^\text{hard}$ in
hard jets with true $p_t = S_1 - q_t$.
However because $M_N$ is a slowly varying function of $p_t$, by
taking the result as contributing to $M_N^\text{hard}(S_1)$ rather than
$M_N^\text{hard}(S_1-q_t)$ we make only a small mistake, of the same order as
other terms that we shall neglect.\footnote{This can be seen by
  observing that $M_N$ satisfies a DGLAP-style equation for its
  evolution $dM_N/d\ln p_t \sim \alpha_s M_N$. Given that $q_t$ is
  itself small compared to $p_t$, $M_N$ for jets with $p_t = S_1 -
  q_t$ differs from that for jets with $p_t = S_1$ by a relative
  amount $\sim \alpha_s q_t/S_1$, which we can neglect in the same way
  that we neglect $\order{q_t^2}$ terms.
  If we wanted to improve on this approximation, then one approach
  might be, for each event, to use the middle expression in
  Eq.~(\ref{eq:MN-hard-v-qt}) and assign it to $M_N$ at $p_t = S_1 -
  \langle q_t \rangle$.
  We have not, however, investigated this option in detail and other
  improvements would probably also be necessary at a similar accuracy,
  e.g.\ taking into account deviations from the simple exponential and
  Gaussian approximations that we have used for $H(p_t)$ and $B(q_t)$.
}

Retaining the terms linear in $q_t$ in Eq.~(\ref{eq:MN-hard-v-qt}) and
averaging now over possible $q_t$ values, making use of
Eq.~(\ref{eq:av-qt}), leads to the following prescription for an
``improved'' subtracted $M_N(S_1)$, corrected for fluctuations effects 
up to first order in $q_t/S_1$:
\begin{equation}
  \label{eq:subimp-v3}
  M_N^\text{sub,imp} =  M_N^\text{sub} \times \left(1 +
    N \frac{\sigma^2 A}{S_1 \mu}\right)
   - r_N\frac{ \sigma \sigma_N A}{\mu S_1^N}\,.
\end{equation}
This is simpler than the corresponding correction would be
directly in $z$ space, in particular because in $z$ space the
correction to one bin of the fragmentation function depends in a
non-trivial way on the contents of nearby bins.
One can think of the advantages of moment space as being that the
correction to a given $N$ value does not depend on $M_N$ at all other
values of $N$, and that it is straightforward to account for
correlations between fluctuations in the jet-$p_t$ and in the moments.

Note that in a real experimental context, calorimeter fluctuations of
the reconstructed jet and background $p_t$'s would have an effect akin
to increasing $\sigma$ and decreasing the correlation coefficient
$r_N$.
The in-situ methods that we use for the determination of $\sigma$,
$\sigma_N$ and $r_N$ would automatically take this into account.
Noise-reduction methods in the reconstruction of the jet $p_t$, as
used by CMS~\cite{Chatrchyan:2011sx}, would have the effect of
reducing $\sigma$ (and probably also $r_N$).
However noise reduction is likely to complicate the meaningful
determination of $r_N$, since it acts differently on pure background
jets as compared to jets with a hard fragmenting component.

\begin{figure}[tp]
  \centering
  \includegraphics[angle=270,width=0.7\textwidth]{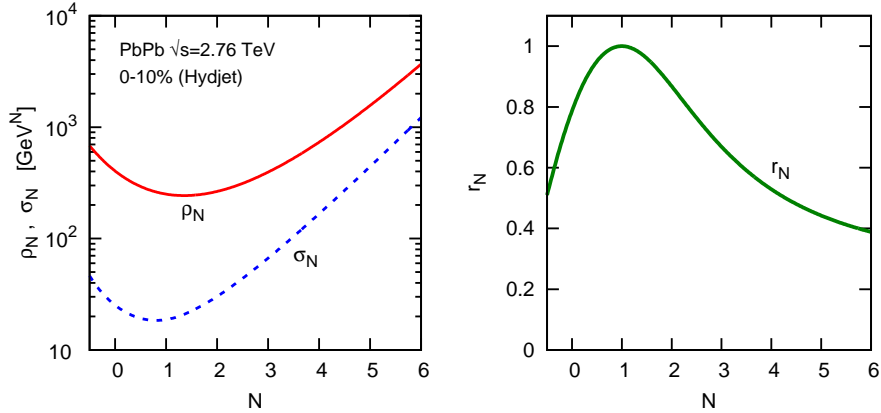}
  \caption{The quantities $\rho_N$, $\sigma_N$ and the correlation
    coefficient $r_N$, shown as a function of $N$ for $0-10\%$ central
    Pb Pb collisions $\sqrt{s_{NN}} = 2.76 \TeV$ as obtained from
      simulations with Hydjet.}
  \label{fig:correlNrhoNsigmaN}
\end{figure}

The correction in Eq.~(\ref{eq:subimp-v3}) can be applied jet-by-jet to
correct for the fluctuation effects. It requires the prior knowledge
of the slope $\mu$ of the jet cross-section, which can be obtained
from $pp$ data, or from simulations.\footnote{In practice, $\mu$
  depends on $p_t$ and should be taken at the scale $S_1-q_t$ in the
  integrand. However, $\mu$ varies slowly with $p_t$ and can easily be
  taken at the fixed scale $p_t$ in our small-fluctuations limit. In
  our analysis $\mu$ ranged from $\sim 9$ GeV at $p_t \simeq 50$~GeV
  to $\sim 28$~GeV at $p_t \simeq 200$~GeV. At $p_t \simeq 100$~GeV we
  had $\mu \sim 16$~GeV.} All the other ingredients that enter this
equation ($\sigma$, $A$, $S_N$, $\sigma_N$, $r_N$) can instead be
determined event-by event or jet-by-jet.
In practice we determine $\sigma$, $\sigma_N$ and $r_N$ from the
ensemble of jets contained in an annulus (or ``doughnut'') of outer
radius $3R$ and inner radius $R$, centred on the jet of interest.
Typical values of $\rho_N$, $\sigma_N$ and $r_N$ are presented as a
function of $N$ in Fig.~\ref{fig:correlNrhoNsigmaN}.

\begin{figure}[tp]
\centerline{\includegraphics[angle=270,width=0.8\textwidth]{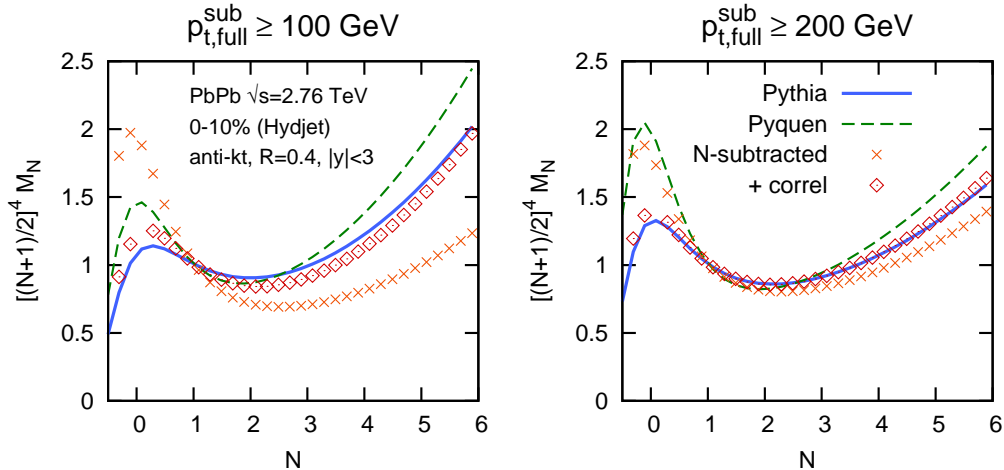}}
\caption{ %
  Jet fragmentation function moments, showing the plain Pythia result,
  the result after embedding in Hydjet and applying plain subtraction
  moment-space subtraction (``N-subtracted'') and after the additional
  improvement to account for correlations (``+ correl''),
  Eq.~(\ref{eq:subimp-v3}).  %
  A quenched result (``Pyquen'') is also shown, to help give an
  indication of the order of magnitude of quenching effects as
  compared to residual misreconstruction effects.
}\label{fig:hydjet+correl}
\end{figure}

We show in figure \ref{fig:hydjet+correl} the result of applying
Eq.~(\ref{eq:subimp-v3}) to our subtraction in moment space, for two
jet $p_t$ thresholds. 
The solid blue curve ($pp$ reference) and orange crosses
(N-subtracted) are identical to the results in the rightmost plots in
figure~\ref{fig:hydjet}.
In addition, figure \ref{fig:hydjet+correl} also displays results
obtained using Eq.~(\ref{eq:subimp-v3}), shown as red diamonds.
One sees how the quality of the agreement with the `hard' blue curve
is markedly improved.
At low $N$ it is the last, additive, term in 
Eq.~(\ref{eq:subimp-v3}) that dominates this improvement, accounting for
the correlation between a jet's reconstructed $p_t$ fluctuations and
the fluctuations in the background's contribution to the moment;
at high $N$ it is the multiplicative $N\sigma^2 A/S_1\mu$ term  that
dominates, correcting for the
fact that the more common upwards fluctuations in the jet's
reconstructed $p_t$ cause the fragmentation $z$ value to be
underestimated.

Figure \ref{fig:hydjet+correl} also shows (green dashed curve) the jet
fragmentation function predicted by the quenching model used in
Pyquen. 
One observes that the remaining deficiencies of the reconstruction
are significantly smaller than the difference between the unquenched
(solid blue) and the quenched (dashed green) FFs, pointing to a
potential discriminating power.
This is to be contrasted with subtraction without improvement (orange
crosses), which especially in the soft region, $N<1$, fails to describe
the blue curve sufficiently well to tell whether quenching (as
predicted by Pyquen) is present or simply that an imperfect
reconstruction is taking place.
This serves as an illustration that the improved subtraction
may now be sufficiently good to allow one to discriminate a quenched
FF from an unquenched one.

We conclude this section by noting that an implementation of the tools
needed to implement the jet fragmentation function subtraction in
moments space, as well as the fluctuations unfolding improvement, will
be made available as a FastJet add-on from
\url{http://fastjet.hepforge.org/contrib}.

%
\section{Conclusions}\label{sec:ccl}

In this paper we have suggested that it could be of interest to study
jet fragmentation-function moments in heavy-ion collisions. 
Individual moments contain specific information such as jet-hadron
multiplicities, the momentum fraction carried by charged particle and
can also provide insight into the relation between inclusive hadron
and inclusive spectra.

As we have seen, it is quite straightforward to correct moments
for effects of fluctuating backgrounds, arguably more so than for
fragmentation functions expressed in terms of momentum fractions.
The correction procedure we discussed is amenable to systematic
improvement, for example by accounting more completely for the shape of
the jet $p_t$ spectrum.
Such extensions are best considered as part of a more detailed
analysis, for example in the context of a full experimental study.

\section*{Acknowledgements}

This work was in part supported by the French Agence Nationale de la
Recherche, under grants ANR-09-BLAN-0060 and ANR-10-CEXC-009-01 and by
the EU ITN grant LHCPhenoNet, PITN-GA-2010-264564.
GPS wishes to thank Guilherme Milhano for related discussions. MC wishes to
thank Peter Jacobs for useful conversations and comments on the draft.

\end{document}